\documentclass[iop]{emulateapj}

\usepackage{mathrsfs }
\usepackage{natbib}
\usepackage{amssymb}
\usepackage{amsmath}
\usepackage{ulem} 

\usepackage{graphicx}

\newcommand{\msun}{$M_{\odot}$}

\begin{document}

\title{Understanding large-scale structure in the SSA22 protocluster region using cosmological simulations\altaffilmark{1}}

\author{
 Michael W. Topping,\altaffilmark{2}
 Alice E. Shapley,\altaffilmark{2}
 Charles C. Steidel,\altaffilmark{3}
 Smadar Naoz,\altaffilmark{2}
 Joel R. Primack\altaffilmark{4}
 }

\altaffiltext{1}{Based on data obtained at the W.M. Keck Observatory, which is operated as a scientific partnership among the California Institute of Technology, the University of California,  and the National Aeronautics and Space Administration, and was made possible by the generous financial support of the W.M. Keck Foundation.}
\altaffiltext{2}{Department of Physics and Astronomy, University of California, Los Angeles, 430 Portola Plaza, Los Angeles, CA 90095, USA}
\altaffiltext{3}{Cahill Center for Astronomy and Astrophysics, California Institute of Technology, 1216 East California Boulevard., MS 249-17, Pasadena, CA 91125, USA}
\altaffiltext{4}{Physics Department, University of California, Santa Cruz, CA 95064, USA}

\email{mtopping@astro.ucla.edu}

\shortauthors{Topping et al.}


\shorttitle{Understanding large-scale structure in the SSA22 protocluster region using cosmological simulations}


\begin{abstract}
We investigate the nature and evolution of large-scale structure within
the SSA22 protocluster region at $z=3.09$ using cosmological
simulations. A redshift histogram constructed from
current spectroscopic observations of the SSA22 protocluster
reveals two separate peaks at $z = 3.065$ (blue) and $z = 3.095$ (red).
Based on these data, we report updated overdensity and mass
calculations for the SSA22 protocluster. We find
$\delta_{b,gal}=4.8 \pm 1.8$, $\delta_{r,gal}=9.5 \pm 2.0$
for the blue and red peaks, respectively, and $\delta_{t,gal}=7.6\pm 1.4$
for the entire region. These overdensities correspond to masses
of $M_b = (0.76 \pm 0.17) \times  10^{15} h^{-1}
M_{\odot}$,  $M_r = (2.15 \pm 0.32) \times  10^{15} h^{-1}
M_{\odot}$, and $M_t=(3.19 \pm 0.40) \times  10^{15} h^{-1}
M_{\odot}$ for the red, blue, and total peaks, respectively.
We use the Small MultiDark Planck (SMDPL) simulation
to identify comparably massive $z\sim 3$ protoclusters,
and uncover the underlying structure and ultimate
fate of the SSA22 protocluster.
For this analysis, we construct mock redshift histograms
for each simulated $z\sim 3$ protocluster, quantitatively comparing
them with the observed SSA22 data. We find
that the observed double-peaked structure in the SSA22 redshift
histogram corresponds not to a single coalescing cluster, but rather
the proximity of a $\sim 10^{15}h^{-1} M_{\odot}$ protocluster
and at least one $>10^{14} h^{-1} M_{\odot}$ cluster progenitor.
Such associations in the SMDPL simulation
are easily understood within the framework of hierarchical clustering
of dark matter halos. We finally find that the opportunity to
observe such a phenomenon is incredibly rare, with an occurrence
rate of $7.4h^3 \mbox{ Gpc}^{-3}$.
\end{abstract}



\section{Introduction}
\label{sec:introduction}
As the largest gravitationally bound structures, galaxy clusters are ideal objects for probing the formation of large scale structure in the universe. Due to their extreme nature, galaxy clusters and protoclusters are an optimal setting to study the effects of environment on galaxy formation and evolution.  The progenitors of todays galaxy clusters, i.e. ``protoclusters" have been identified all the way out to $z\sim6$, using a variety of techniques \citep{Toshikawa2014}.    The study of galaxy clusters and protoclusters is further aided by the multiple techniques that have been developed in order to find them.

There are currently many techniques for finding high-redshift protoclusters, including the serendipitous identification of redshift overdensities within spectroscopic surveys of Lyman Break Galaxies (LBGs), Ly$\alpha$ emitters (LAEs) or other magnitude-limited galaxy samples \citep{Steidel1998, Steidel2003, Steidel2005, Harikane2017, Chiang2015, Lemaux2014}, targeted searches for LAEs around radio galaxies \citep[e.g.][]{Venemans2007}, and Ly$\alpha$ forest tomography \citep{Lee2016}. Based on these several methods, the number of known  $z>2$ protoclusters has grown dramatically over the past decade. Studying the key high-redshift epoch of structure formation, when the clusters are still collapsing, helps to give us a more complete picture of massive galaxy clusters and their environments at $z=0$.

\citet{Steidel1998} reported the discovery of the SSA22 galaxy protocluster at $z=3.09$ within a large survey of $z\sim3$ LBGs, and measured an overdensity of $\delta_{gal}=3$, with the expectation of the overdensity evolving into a massive Coma-like cluster with a mass of $M\sim10^{15}M_{\odot}$ by $z=0$. Based on an expanded dataset, \citet{Steidel2000} obtained a revised estimate for the overdensity of $\delta_{gal}=6.0\pm 1.2$.  Since then, the area surrounding the $z\sim3.09$ overdensity has been observed through multiple observing campaigns spanning from radio to X-ray wavelengths.  These studies have revealed tens of Lyman alpha blobs \citep{Matsuda2011, Geach2005, Geach2016}, and multiple X-ray sources \citep{Lehmer2009, Geach2009}.  Additional studies include deep ALMA observations in the central region of the protocluster \citep{Umehata2015, Geach2016, Hayatsu2017}, near-infrared spectroscopic observations of massive red K-band-selected galaxies \citep{Kubo2015}, and high-resolution \textit{Hubble Space Telescope} imaging \citep{Chapman2004}.

In addition to the extensive multi-wavelength studies of SSA22, followup spectroscopic observations have revealed details about structure within the overdensity.  \citet{Matsuda2005} mapped the three-dimensional structure of LAEs in and around the protocluster, and reported  evidence for large-scale filamentary structure. \citet{Topping2016} showed the existence of two distinct groups of galaxies, both LAEs and LBGs, separated both on the sky and in redshift space, and observed as a double-peaked redshift histogram.  This structure was discovered by focusing on the highest density region of the protocluster, but remains persistent when the observed region is expanded \citep{Topping2016, Yamada2012}.  From these studies it is unclear what the evolution and fate of the $z\sim3.09$ protocluster and its surrounding structure will be down to $z=0$. In particular, we would like to understand if these structures will coalesce, or remain distinct throughout their evolution.

Cosmological N-body simulations provide a useful tool for studying the evolution of large scale structure.  Recently, the increase in computational power leads to cosmological simulations with both higher resolution extending down to lower-mass halos, and larger volumes including the largest, rarest structures in the universe.  These advances, combined with the availability of easily searchable halo catalogs and merger trees, enable us to use simulations to understand the underlying physical structures observed in SSA22, and how they evolve to the present day.

In this paper, we further investigate the nature of the large-scale structure presented by \citet{Topping2016}.  We utilize the halo catalogs and merger tree information from the Small MultiDark Planck (SMDPL) dark matter simulation \citep{Rodriguez-Puebla2016, Klypin2016, Behroozi2013a}, which has sufficient resolution and simulation volume to compare multiple simulated protoclusters with our observations. We examine massive overdensities at the redshift of the SSA22 protocluster in order to understand the intrinsic physical structure giving rise to the observed structure at $z\sim3$, and what such structure evolves into by $z=0$. Section~\ref{sec:SSA22-field} describes our observations and the calculation of an updated overdensity and mass estimate for the SSA22 protocluster based on current spectroscopic data.  Section~\ref{sec:methods} describes the cosmological simulation used to interpret the SSA22 observations, and the methods used to compare it to the observations.  Section~\ref{sec:results} shows the results of a comparison between the observations and simulations. Finally, Section~\ref{sec:discussion} discusses an analytic approach to understanding the results from the simulation, and a calculation of the cosmic abundance of large-scale structure similar to the observed structure in SSA22.  This paper adopts a cosmology of $\Omega_m=0.3$, $\Omega_{\Lambda}=0.7$, $n_s = 0.96$, $\sigma_8 = 0.8228$, and $h=1.0$, unless otherwise stated.  We also use the abbreviation cMpc for ``comoving Mpc."

\section{SSA22 Field}
\label{sec:SSA22-field}

\subsection{Data}
\label{sec:data}
Our sample consists of LBGs and LAEs with spectroscopic measurements within a $9^{\prime}\times9^{\prime}$ region of the SSA22 field, centered on R.A.=22:17:34, decl.=00:15:04 $(\rm J2000)$, as described by \citet{Steidel1998}.  The LBGs in our sample were selected as part of the survey of $z\sim3$ star-forming galaxies presented in \citet{Steidel2003}. The LAEs were first identified using broadband BV imaging from Subaru/Suprime-cam in addition to narrowband imaging from Keck/LRIS and Subaru/Suprime-cam using a filter tuned to the wavelength of Ly$\alpha$ at $z=3.09$ (centered on 4985\AA\  with a bandwidth of 80\AA). The LAEs were selected based on BV$-$NB4985 colors indicating a narrowband excess, which ensures a sample of galaxies with large ($>20$\AA) Ly$\alpha$ EWs at redshifts coincident with the central density peak of SSA22 ($3.05 \lesssim z \lesssim 3.12$).  The spectroscopic measurements for galaxies in the SSA22 field were obtained using the LRIS spectrograph at the Keck observatory across multiple observing campaigns and instrumental configurations \citep{Steidel2003, Nestor2011, Nestor2013}.  A more detailed description of the redshift determinations can be found in \citet{Topping2016}, and further details about the observations and data reduction can be found in \citet{Steidel2003} and \citet{Nestor2011, Nestor2013}.

\begin{figure}[h]
\includegraphics[width=1\linewidth]{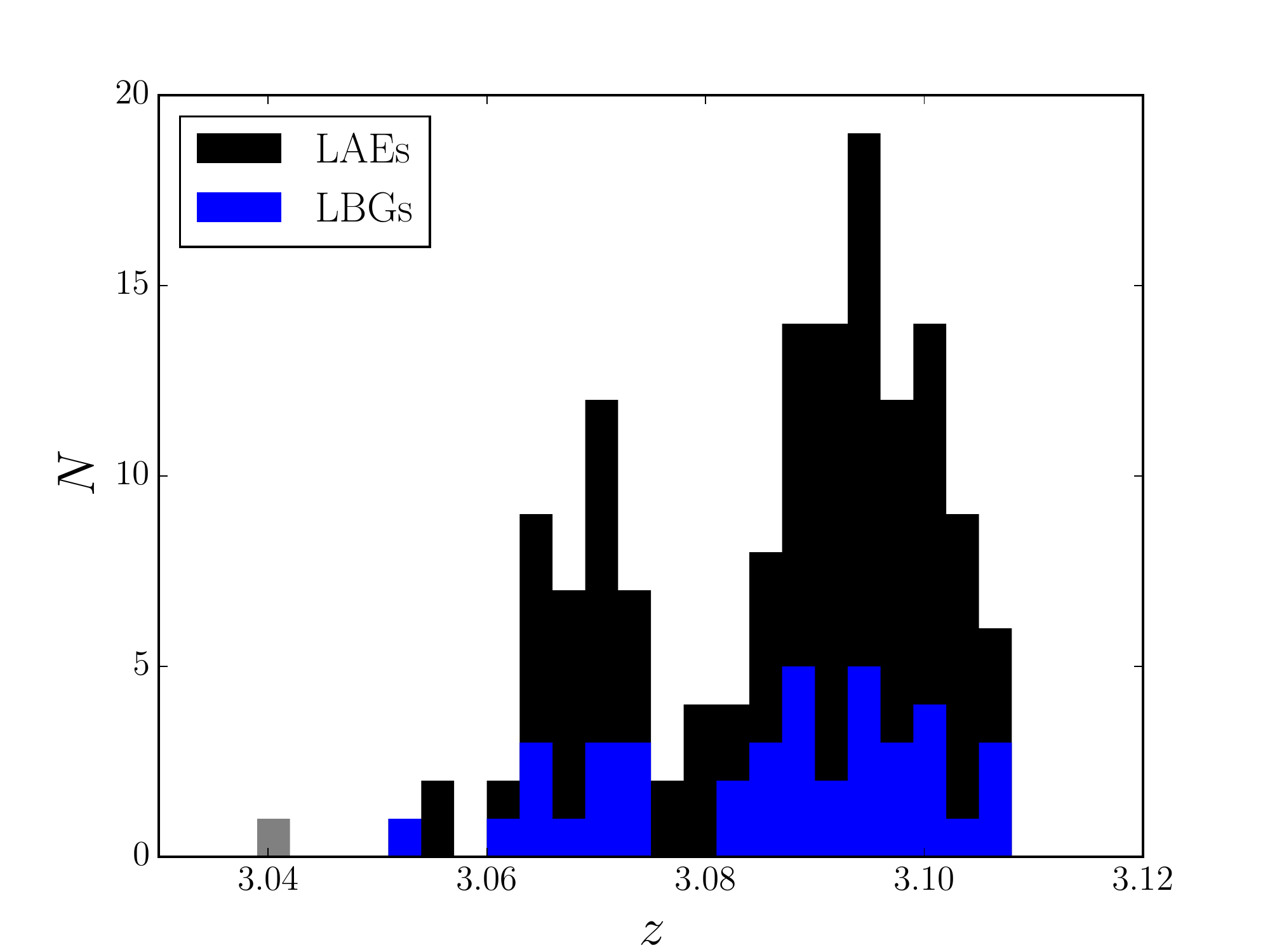}
\caption{Redshift histogram of LAEs and LBGs in the SSA22 field.  The double peaked morphology is clearly present with peaks at $z=3.069$ (blue peak) and $z=3.095$ (red peak).  The blue histogram shows the contribution from the LBGs, and the remaining black histogram is the contribution from LAEs.}
\label{fig:histogram}
\end{figure}

We determined the systemic redshift of galaxies in the SSA22 field by measuring the redshift of Ly$\alpha$ emission, interstellar metal absorption lines, or both, and removing the effects of large-scale gas outflows.  We applied the formulas presented in \citet{Trainor2015} for LAEs and \citet{Adelberger2003} for LBGs, to translate from the observed rest-UV emission and absorption redshifts to the true, systemic redshifts.  We compiled the resulting systemic redshifts of galaxies within SSA22 into a redshift histogram (Figure~\ref{fig:histogram}).  Galaxies in the SSA22 redshift histogram are clearly separated into peaks centered at $z=3.069$ (blue) and $z=3.095$ (red) with widths $\sigma_{z,b}=0.0047$ and $\sigma_{z,r}=0.0074$ respectively. Hereafter, we describe the total, blue, and red regions using the subscripts $t$, $b$, and $r$ respectively.

\subsection{Galaxy Overdensity Calculation}
\label{sec:deltagal}
\begin{figure}[h]
\includegraphics[width=1\linewidth]{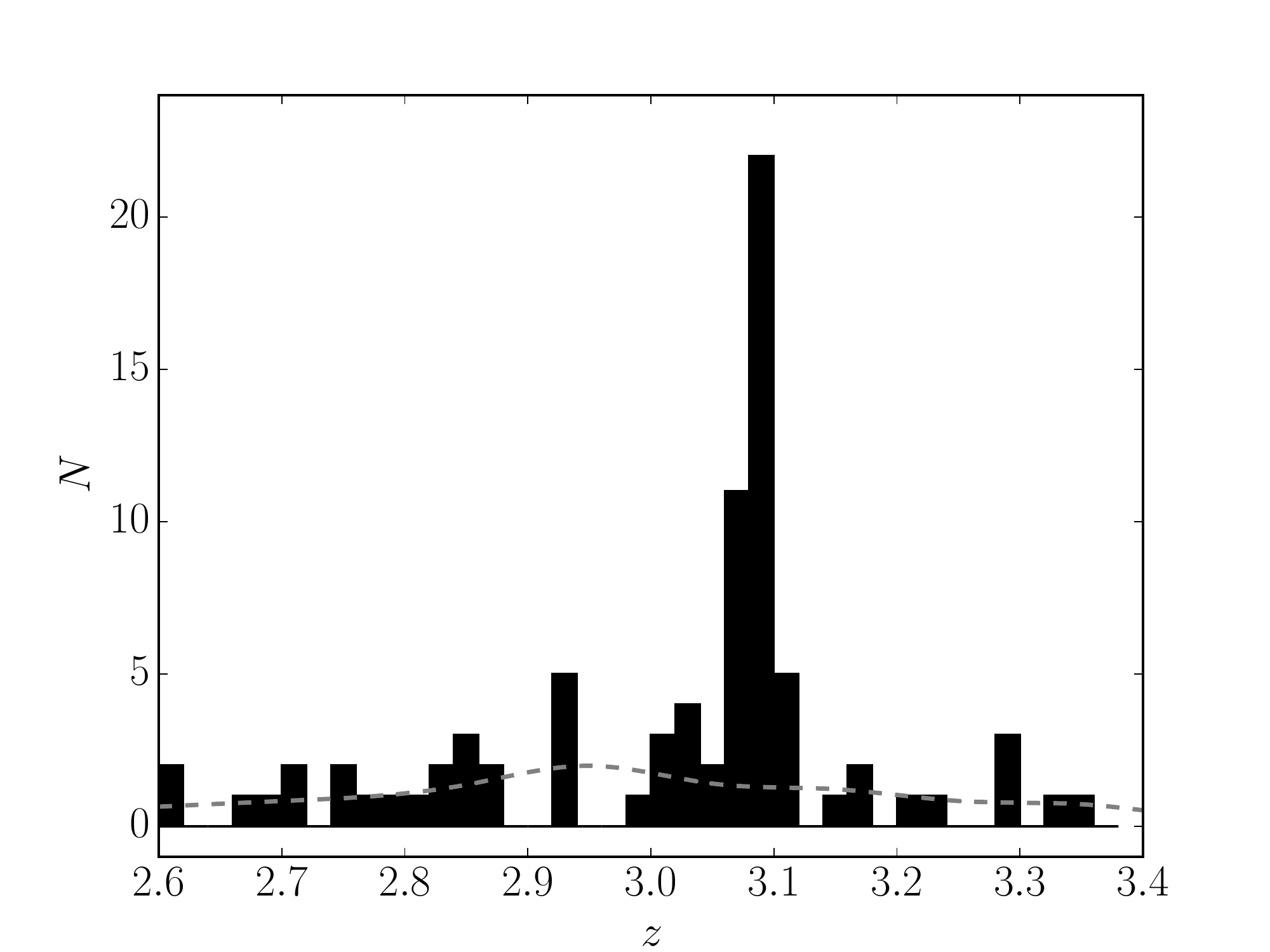}
\caption{Redshift histogram of LBGs observed in the SSA22 field.  The grey dashed line shows the LBG selection function determined using $883$ LBGs, and normalized using the method described in the text. The bin size used in this redshift histogram is too coarse to observe the double-peaked structure near $z=3.09$.}
\label{fig:sfunction}
\end{figure}

The significance of the SSA22 overdensity has been calculated in past work \citep{Steidel1998, Steidel2000}. However, given our significantly larger sample of spectroscopic redshifts in SSA22 \citep{Topping2016}, and the updated LBG redshift selection function \citep{Steidel2003}, it is worth revisiting this calculation. To estimate the galaxy overdensity qualitatively, we compared the number of galaxies contained in the SSA22 redshift spike $(N_{\rm obs})$ with the number of galaxies expected in the same redshift interval from the LBG average selection function $(N_{\rm expect})$.  For this calculation, we restricted $N_{\rm obs}$ to the LBGs in our observed sample and did not consider LAEs, since the LBGs have a well-defined redshift selection function.  We define the galaxy overdensity, $\delta_{gal}$, as:
\begin{equation}
\label{eqn:dgal}
\delta_{gal} = \frac{N_{\rm obs}}{N_{\rm expect}} - 1 .
\end{equation}
The observed sample used for this calculation includes 82 LBGs in the redshift interval $2.6 \le z \le 3.4$.  The redshift histogram of these galaxies is shown in Figure~\ref{fig:sfunction}, where the well-known overdensity at $z\sim3.09$ is clearly visible.

To construct the LBG selection function, we used the sample of LBGs from \citet{Steidel2003}, with one key difference. The inclusion of SSA22 galaxies in the sample would increase the value of the selection function within the $z=3.09$ spike interval, thus biasing the inferred overdensity towards lower values. Therefore, we excluded these galaxies, with $883$ redshifts remaining.  We fit a spline to the histogram of the remaining galaxies, which resulted in a smooth selection function. Finally, we normalized the selection to the SSA22 redshift histogram, which allowed us to directly compare the number of LBGs in a given redshift interval. Determining the correct normalization is a key step in calculating the galaxy overdensity.  Specifically, we normalized the LBG selection function such that its integral over the redshift ranges $2.6 \le z \le 3.03$ and $3.12 \le z \le 3.4$, was equal to the number of observed galaxies in the SSA22 field in the same redshift intervals.  These ranges were chosen to match the number of ``field" galaxies in SSA22 and the overall LBG selection function.  The resulting selection function is displayed in Figure~\ref{fig:sfunction} overlaid on the SSA22 LBG redshift histogram.

Using the redshift histogram and the LBG selection function, we computed the galaxy overdensity of SSA22.  In detail, we calculated the galaxy overdensity for three components of the SSA22 protocluster: the blue peak, the red peak, and the total volume.  We carefully determined the boundaries of the redshift intervals in order to accurately calculate the overdensity.  In contrast to previous work, here we found that the low and high redshift boundaries of the total SSA22 interval were self-evident, as defined by a large gap on either side of the redshift distribution, with the boundaries occurring at the redshift of the last galaxy on each side of the overdensity.  Therefore, we set the low and high redshift boundaries to $z=3.0598$, and $z=3.1048$ respectively, and removed the galaxies that define these boundaries from our future calculations.  To find the boundary that separates the red and blue peaks, we fit the sum of two Gaussians to the redshift histogram, and determined the redshift at the minimum of the trough between the two peaks.  We measured this boundary to be at $z=3.0788$.

Due to the effects of redshift-space distortions \citep{Kaiser1987}, and the fact that the SSA22 protocluster is collapsing, the redshift intervals we defined are contracted compared to the ranges defined by the physical size of the protocluster in the Hubble flow.  We used a correction factor $C$ \citep{Padmanabhan1993} to quantify this effect, as defined by:
\begin{equation}
\label{eqn:C}
C = 1+f-f(1-\delta_m)^{\frac{1}{3}},
\end{equation}
where
\begin{equation}
f=\frac{d\ln{D}}{d\ln{a}},
\end{equation}
$D$ is the linear growth factor, $a$ is the cosmological scale factor, and $\delta_m$ is the matter overdensity, related to $\delta_{gal}$ through:
\begin{equation}
\label{eqn:deltam}
1+b\delta_m = C(1+\delta_{gal}).
\end{equation}
We defined the LBG bias factor, $b$ (Equation~\ref{eqn:bias}), by comparing $\sigma_{8,gal}$, the LBG number fluctuations, and $\sigma_{8,CDM} = 0.8228$, which corresponds to $\sigma_{8,CDM}\rvert_{z=3.09} = 0.254$ at $z=3.09$ \citep{Planck2013}.
\begin{equation}
\label{eqn:bias}
b^2 = \frac{\sigma_{8,gal}^2}{\sigma_{8,CDM}^2}\Biggr\rvert_{z=3.09}
\end{equation}
We calculated the value of $\sigma_{8,gal}$ using the correlation length, $r_0$, and the slope, $\gamma$, from the LBG autocorrelation function:
\begin{equation}
\label{eqn:sigma8}
\sigma_{8,gal} = \frac{72(\frac{r_0}{8\rm \ cMpc})^{\gamma}}{2^{\gamma}(3-\gamma)(4-\gamma)(6-\gamma)}
\end{equation}
\citep{Peebles1980}. For these calculations, we found a value of $f=0.986$, and adopted values of $r_0=6.0\pm0.5 h^{-1} \rm \ Mpc$ and $\gamma=1.5$ from \citet{Trainor2012}, which result in a bias of $b=3.84\pm0.25$. We estimated the errors of the bias from the uncertainties of the autocorrelation function parameters, $r_0$ and $\gamma$, and $\sigma_{8,CDM}$.  Table~\ref{table:masscalc} shows the values of these parameters resulting from our calculation.

Neglecting the effects of large-scale redshift-space distortions (i.e., infall) in estimating the number of LBGs expected from the LBG selection function causes us to underestimate the relevant redshift interval, and therefore the expected number of galaxies, $N_{expect}$, as well. We corrected for this effect by increasing the redshift interval by a factor of $1/C$ (see Table~\ref{table:masscalc}) when integrating the LBG selection function, and recalculating the number of galaxies expected within the interval, as well as the associated galaxy overdensity.  One subtlety lies in the fact that our correction factor, $C$, was initially calculated based on an overdensity that was overestimated due to the underestimate of the selection function, resulting in a correction that is too large.  We therefore recomputed the correction factor using the updated overdensity, and repeated the procedure of correcting the redshift interval of the selection function, and recalculating the overdensity.  We iterated this process until the galaxy overdensity converged to its true value, which we adopted as our final value for the overdensity.  We obtained overdensities of $\delta_{t,gal}=7.6\pm1.4$, $\delta_{r,gal}=9.5\pm 2.0$, and $\delta_{b,gal}=4.8 \pm 1.8$, for the total, red, and blue regions respectively.  Our updated total overdensity is larger than the value previously reported in \citet{Steidel1998} ($\delta_{t,gal}=3.6^{+1.4}_{-1.2}$) but consistent with the value reported in \citet{Steidel2000} ($\delta_{t,gal}=6.0\pm1.2$).

\subsection{Mass Calculation}
\label{sec:masscalc}
Using the updated estimates of the galaxy overdensity and appropriate volume for each section of the protocluster, corrected for the effects of redshift distortion, we computed the total, blue-peak and red-peak protocluster masses using:
\begin{equation}
\label{eqn:Mass}
M = \bar{\rho}V_{\rm true}(1+\delta_m),
\end{equation}
where $\bar{\rho}$ is the mean density of the universe, and $V_{\rm true}=V_{\rm apparent}/C$.

We calculated the mass overdensity, $\delta_m$, of each region using Equation~\ref{eqn:deltam}, utilizing the values for the correction factors, $C$ (see Table~\ref{table:masscalc}), that we obtained at the end of the iterative process described above.  Using these correction factors, we calculated mass overdensities of $\delta_{t,m}=1.3\pm0.4$, $\delta_{r,m}=1.5\pm0.4$, and $\delta_{b,m}= 0.9\pm0.3$, for the total cluster, red peak, and blue peak respectively.

In order to estimate $V_{\rm apparent}$ (and the corresponding $V_{\rm true}$) for each region, we multiplied its line-of-sight extent and on-sky area. In the line-of-sight dimension, the spatial extent is represented by the difference in the radial comoving distance between the two redshift boundaries. We used the on-sky coverage of our observations, as described in \citet{Topping2016}, as the area in the transverse dimensions, corresponding to a value of $12 \times 14\ h^{-2} \rm \ cMpc^2$ for the area on the sky.  For the blue peak, we reduced the area on the sky because the galaxies contained within this peak cover only $\sim75\%$ of the observing area \citep{Topping2016}. Our observations, and therefore the area used in our calculations, did not cover the full extent of the protocluster, as probed by e.g., \citet{Matsuda2005} and  \citet{Yamada2012}. Therefore, increasing the volume to enclose the entire protocluster may result in an increased mass estimate.  On the other hand, our observations were centered on the highest density region of the protocluster, so expanding the protocluster volume may dilute the overdensity, therefore negating the expected mass increase caused by using a larger volume.  For example, using the positions presented in \citet{Hayashino2004} we determined that the average surface density of LAEs decreases by $\sim20\%$ if our observing window size is doubled.  Analysis of protocluster membership in the Millennium Simulation shows that only $\sim50\%$ of the galaxies within this area will be gravitationally bound to the main cluster by $z=0$ \citep{Muldrew2015}.  The net result of these two effects is a predicted $z=0$ mass higher than our estimate, but much more uncertain.

Based on the $\delta_m$ and $V_{\rm true}$ values described above, we calculated the mass of the total cluster to be $(3.19\pm 0.40) \times 10^{15}\ h^{-1}\ M_{\odot}$, and calculated the mass of the red (blue) peak to be $(2.15\pm 0.32) \times 10^{15}\ h^{-1}\ M_{\odot}$ ($(0.76 \pm 0.17) \times 10^{15}\ h^{-1}\ M_{\odot}$).  We determined the errors on our mass calculation based on our uncertainties of the mass overdensity.  The volumes encompassing the red and blue peaks do not fill the entire space of the total overdensity, so the sum of the red and blue peak masses is less than the mass of the entire structure.

\begin{deluxetable*}{lccc}
\huge
\centering
\tabletypesize{\footnotesize}
\tablecolumns{4}
\tablecaption{}
\tablehead{
\colhead{} & \colhead{Blue} & \colhead{Red} & \colhead{Total} }
\startdata
$z_{min}$ & 3.0598 & 3.0788 & 3.0598 \\
$z_{max}$ & 3.0788 & 3.1048 & 3.1048 \\
$z_{peak}$ &   $3.069 \pm 0.001$    &  $3.095 \pm 0.001$     & $-$ \\
$N_{expect}$ & 1.71 & 2.66 & 4.40 \\
$N_{obs}$ & 10 & 28 & 38 \\
$\delta_{gal}$ & $4.83 \pm 1.84$ & $9.51\pm 1.99$ & $7.64\pm1.40$ \\
$\delta_{m}$ & 0.9 & 1.509 & 1.285 \\
$C$ & 0.765 & 0.647 & 0.688 \\
Dimensions $[h^{-3}\ \rm cMpc\times cMpc \times cMpc] $ & $\frac{3}{4}\times 12\times 14 \times 18.38$ & $12\times 14\times 24.95$ & $12\times 14\times 43.338$ \\
$V \ [h^{-3}\ \rm cMpc^3]$ & 2315.9 & 4191.6 & 7280.8\\
$M$ & $(0.757 \pm 0.171) \times 10^{15} h^{-1} M_{\odot}$ & $(2.146\pm 0.324) \times 10^{15} h^{-1} M_{\odot}$ & $(3.194\pm 0.401) \times 10^{15} h^{-1}M_{\odot}$ \\
\enddata
\label{table:masscalc}
\end{deluxetable*}

\section{Methods and Simulations}
\label{sec:methods}
We use cosmological N-body simulations in order to better understand the underlying physical structures giving rise to the observed properties of the SSA22 protocluster, as well as its evolution in the context of structure formation.  In this section we present a description of the simulations we used, our technique for identifying protoclusters, and finally the methods that we used to search for analogs of the observed SSA22 structures in the simulation.

\subsection{SMDPL Description}
\label{sec:smdpl}
We use halo catalog and merger tree information drawn from the Small MultiDark Planck (SMDPL) simulation data set\footnote{http://hipacc.ucsc.edu/Bolshoi/MergerTrees.html} \citep{Klypin2016} in order to compare the observed structure in SSA22 to what is found in cosmological N-body simulations \citep{Klypin2016, Behroozi2013a, Behroozi2013b, Rodriguez-Puebla2016}. We chose this simulation because its box size ($400\rm \ h^{-1} \ Mpc$) allows for a large enough sample ($N=19$) of clusters that are within the estimated  $3\sigma$ uncertainty of the mass of the red peak in SSA22 (i.e., $10^{15}\ h^{-1}\ M_{\odot}\le M \le 1.7 \times 10^{15}\ h^{-1}\ M_{\odot}$). Hereafter, we describe masses of halos using their virial mass, $M_{\rm vir}$, defined by \citet{Rodriguez-Puebla2016}. The SMDPL simulation is also characterized by the following cosmological parameters: $\Omega_{m} = 0.307$, $\Omega_{\Lambda}=0.693$, $h=0.678$, $n_s=0.96$, and $\sigma_8=0.829$. These parameters are consistent with current Planck results \citep{Planck2013}, as opposed to those adopted for the Millennium simulation \citep[][$\Omega_{m} = 0.25$, $\Omega_{\Lambda}=0.75$, $h=0.73$, $\sigma_8=0.9$]{Springel2005}. In addition, with a particle mass of $M_{part} = 9.63\times 10^{7}\ h^{-1}$\msun, the mass resolution of the SMDPL simulation allows us to identify robust halos down to the mass that may host galaxies similar to ones in our observations ($M\sim 10^{10.6}\ h^{-1} $\msun).  The halo catalogs are saved in a series of $117$ snapshots, starting at Snapshot Number 0 (called $snapnum$ in the catalogs) at $z=18.56$, and ending with $snapnum=116$ at $z=0$.  The snapshots are saved with a time resolution of $\Delta z\approx 0.16$ at $z\sim3$.  This time resolution allows us to perform our analysis on halos at the epoch of the SSA22 protocluster observations.  The difference between the cosmological parameters used in the SMDPL simulation and our analysis in Section~\ref{sec:SSA22-field} is not significant, and therefore our inferences based on the results are valid.

\subsection{Protocluster Identification}
\label{sec:protocluster-ident}
Based on the mass calculations presented in Section~\ref{sec:masscalc}, we expect the SSA22 protocluster to evolve into a massive ($M\sim 10^{15}\ h^{-1}M_{\odot}$) cluster at $z=0$, so we start by selecting all $z=0$ halos, determined using the ROCKSTAR spherical overdensity method \citep{Behroozi2013b}, in the simulation with masses $M>10^{15}\ h^{-1}$\msun\ from the SMDPL halo catalog.  We identify $19$ systems that meet this criterion. After identifying these halos, we follow their histories through the merger trees constructed from the simulation \citep{Behroozi2013a}, in order to select the progenitor halos at a given epoch ($z=3.03$, $snapnum=31$).  We chose this snapshot as it has the closest redshift to that of the SSA22 protocluster.

\subsection{Methods for Comparison}
\label{sec:comparisonMethods}
We present two methods to search for SSA22 analogs in the SMDPL simulation.  First, we start by assuming that the observed structure in SSA22 will collapse into a massive cluster at $z=0$.  To mimic this regime in our analysis of the SMDPL simulation, we limit our sample to halos that collapse to a single massive structure at $z=0$.  We also employ an alternate, complementary approach in which we construct a sample of halos within a volume surrounding each of the $z\sim3$ protoclusters with no requirement on their status as a member of the descendant cluster at $z=0$.  We then identify what kind of structures form from these halos by $z=0$, and compare them to the current predictions for the fate of the SSA22 protocluster.

\subsubsection{Progenitors Only}
\label{sec:progenitors}

We begin by describing the method that selects our parent sample of halos based on their membership in a single massive structure at $z=0$.  In order to compare any structure present in the simulated protoclusters to the structure observed in SSA22, we constructed redshift histograms from the sample of cluster progenitor halos.  We created redshift histograms by viewing each protocluster from multiple sight lines.  By observing through many sight lines we obtained a comprehensive view of each protocluster, and a better chance of detecting any structure that may be present. We expect adjacent sight lines to show similar evidence of structure, and since each sight line is a different random realization of the protocluster, sampling many sight lines allows us to differentiate between real structure and statistical flukes.  For each protocluster we observed 3600 sight lines, each of which is separated by $6^{\circ}$ in the azimuthal $\theta \in [ 0,2 \pi )$ direction, and $3^{\circ}$ in the polar $\phi \in [ 0,\pi )$ direction.

For a given sight line, the simulated redshift histogram consists of calculated redshifts for 146 halos that are progenitors of a particular protocluster.  We chose this number of halos to be the same as the number of galaxies (both LBGs and LAEs) that have spectroscopic redshifts in SSA22.  To select these halos, we first narrowed down the sample based on their projected positions in the protocluster.  We required that selected halos be within the observed area of SSA22, $\sim12\times14\ h^{-2} \rm \ cMpc^2$, centered on the highest density peak.  To choose the 146 halos whose redshifts make up the redshift histogram for a given sight line, we first randomly selected 40 halos out of all cluster progenitor halos with masses above $M>10^{11.55}\ h^{-1}$\msun \citep{Trainor2012}, corresponding to LBGs in our observed SSA22 sample.  We then randomly selected 106 halos from among the remaining cluster progenitor halos with masses $M>10^{10.6}\ h^{-1}$\msun \citep{Gawiser2007}, which represent the LAEs in our simulated redshift histogram.  This selection process typically results in a sample that contains $\sim 10\%$ of the total cluster progenitors.  This analysis assumes that the LBGs and LAEs in our sample are the central galaxies of their host dark matter halos, as opposed to satellites.  The similar number densities and clustering strengths of LBGs and their host halos \citep{Conroy2008, Trainor2012}, in addition to the low halo occupation fraction of LAEs \citep[1-10\%;][]{Gawiser2007}, suggest that this assumption is valid.

To calculate the observed redshift of a halo, we first required its 3D position and velocity, given in the SMDPL halo catalog. We defined the center of the protocluster as the center-of-mass of all cluster progenitor halos, and set the center of each protocluster to be at $z=3.09$. We calculated the redshift of each halo by determining its line-of-sight distance away from the protocluster center, and the corresponding velocity using the Hubble flow.  In addition, we adjusted the estimated redshift to take into account the line-of-sight peculiar velocity, $\Delta v$, of each halo using $\Delta z = \Delta v/c \times (1+z_H)$, where $z_H$ is the redshift of the halo after taking into account the Hubble flow. We then collected the redshifts of all 146 halos into a redshift histogram.

We began by using the two-sample Kolmogorov-Smirnov (KS) test as a metric for comparison between each simulated protocluster redshift distribution and the observed SSA22 distribution. For each KS test, we determined the probability that the two distributions were drawn from the same parent distribution, a $p$-value. We introduced a $p$-value cutoff of $p\ge 0.4$, which distinguished redshift histograms that were well represented by two peaks, and those that presented only a single peak. We determined the value for this cutoff by trial-end-error. We adjusted the cutoff and visually inspected each qualifying histogram and its best fit models to determine at what $p$-value the histograms are typically double peaked. This cutoff allowed us to exclude those redshift histograms from further analysis that did not show similar structure to that in the SSA22 protocluster.

After we determined the existence of structure in a given sight line, we compared the simulated redshift histogram to the one observed in SSA22.  We first fit the sum of two Gaussians to the simulated redshift histogram.  We then required the associated best fit parameters to be comparable to parameters found for the SSA22 redshift histogram.  The requirements for the parameters of the larger ($l$) and smaller ($s$) peaks are as follows:
\begin{equation}
\label{eqn:criteria}
\begin{split}
0.341 \le \frac{N_s}{N_l} \le 0.493 \\
0.004 \le \sigma_l \le 0.01 \\
0.004 \le \sigma_s \le 0.01 \\
0.02 \le \Delta z \le 0.032.
\end{split}
\end{equation}
In these expressions, $N_s$ and $N_l$ are the number of galaxies in the smaller and larger peaks, respectively.  We determined a boundary at the trough between the two peaks, and counted the number of galaxies on either side.  We define $\sigma_l$ and $\sigma_s$ as the best-fit standard deviations, in redshift units, of the large and small peaks, respectively, and $\Delta z$ as the redshift difference between the centers of the two peaks.  The existence of redshift histograms that fit these criteria would suggest that the observed structure in SSA22 may collapse into a single massive cluster at $z=0$.

\subsubsection{Halos in Surrounding Volume}
\label{sec:volume}
In addition to searching for structure within the distribution of the $z\sim3$ progenitors of a single massive $z=0$ cluster, we also investigated halos in a volume surrounding each protocluster, regardless of their membership in a particular $z=0$ structure.  The full width covered by the SSA22 redshift histogram corresponds to a distance of $\sim42\ h^{-1}\rm \ cMpc$ along the line of sight. Accordingly, to isolate a comparable volume in the simulation, we began by selecting all halos within a $42\ h^{-1} \rm \ cMpc$ radius from the center for each of the 19 identified protoclusters.  We then followed the procedure described in Section~\ref{sec:progenitors} of selecting $146$ halos, calculating redshifts, creating redshift histograms, and determining the similarity of the simulated and observed SSA22 redshift histograms, for $3600$ sight lines of each protocluster.

We used the halo merger trees to determine the $z=0$ structures formed from galaxies present in the $z\sim3$ redshift distribution selected in this volume-limited manner.  Accordingly, the underlying nature and evolution of double-peaked structure in a protocluster at $z\sim3$, identified with this method as being analogous to the SSA22 protocluster, will then shed light on the potential fate of the observed structures in SSA22.

\section{Results}
\label{sec:results}

\subsection{Protocluster Members}
\label{progenitors-res}

We first tested the assumption that the double-peaked redshift histogram is representative of the progenitors of a single massive ($M\ge10^{15}\ h^{-1}$\msun) protocluster at $z=0$. Under this assumption, we expect that the majority of the galaxies in SSA22 will collapse into a massive cluster at $z=0$.  By investigating the $z=3$ cluster progenitors of massive clusters at $z=0$, we are able to identify which, if any, parts of the structure will be a component of the cluster once it has collapsed.

\begin{figure*}
    \centering
    \begin{minipage}[t]{0.48\textwidth}
        \centering
        \includegraphics[width=\linewidth]{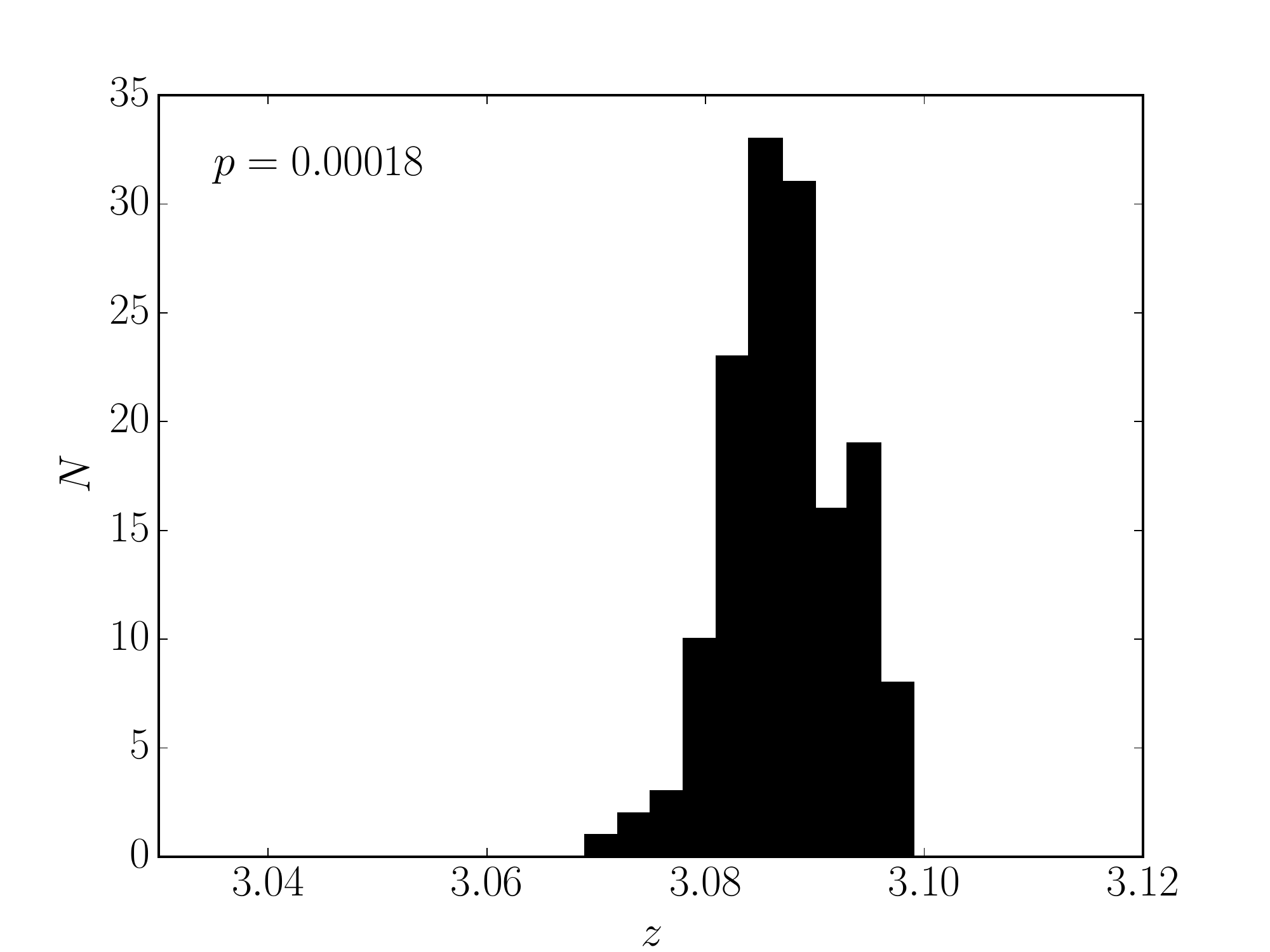}
        \caption{Redshift histogram calculating using only halos that are cluster progenitors.  This is the redshift histogram that is the most similar to that of SSA22 across all sight lines in each of our 19 protoclusters. Even in this case the $p$-value suggests that the observed and simulated redshift distributions are significantly different.}
        \label{fig:prog-hist}
    \end{minipage}
    \hfill
    \begin{minipage}[t]{0.48\textwidth}
        \centering
        \includegraphics[width=\linewidth]{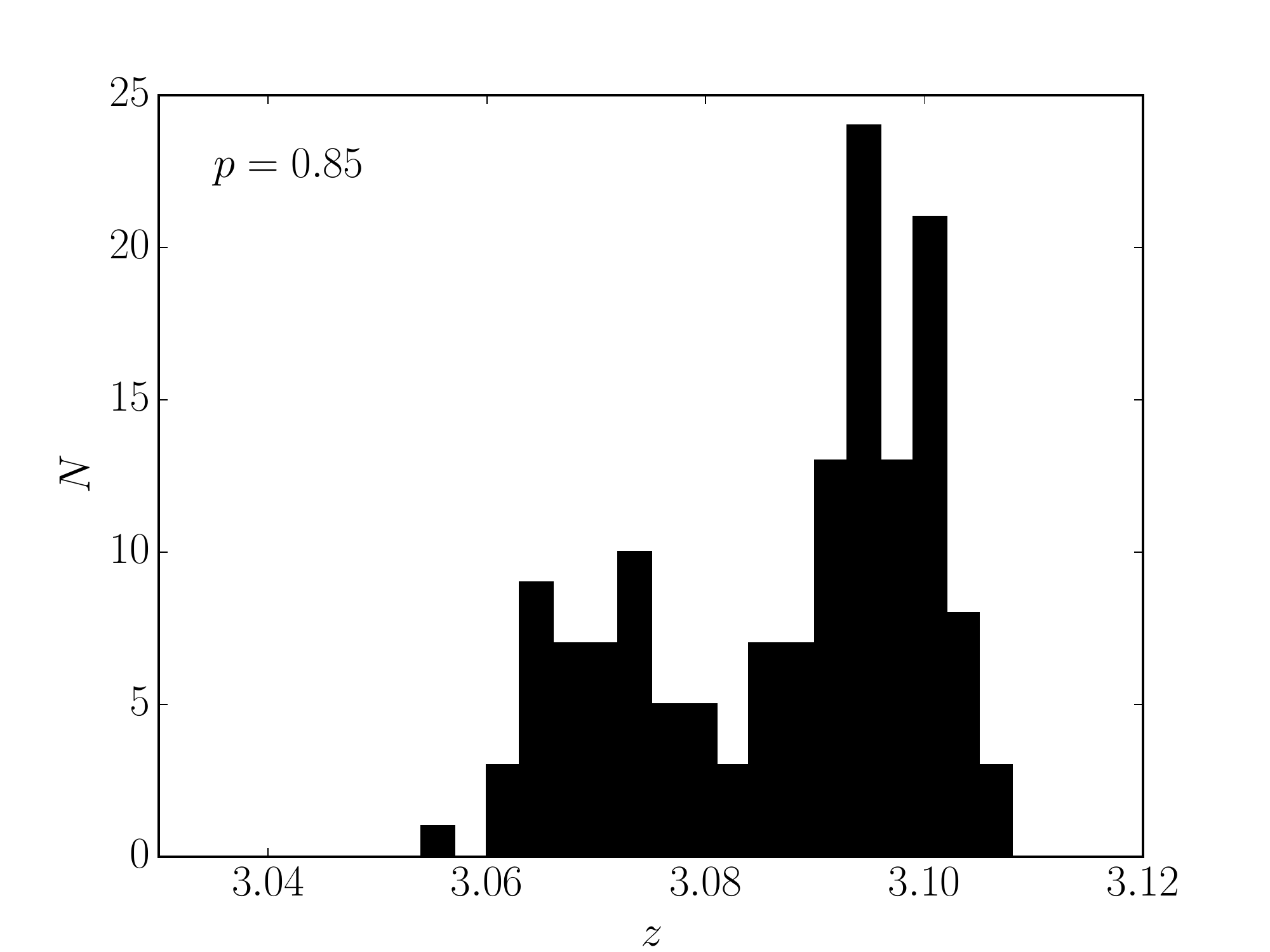}
        \caption{Example of a double-peaked redshift histogram computed by selecting cluster progenitor halos, as well as halos in the volume surrounding the protocluster.  We determined this redshift histogram to fit the SSA22 similarity criteria given in Equation~\ref{eqn:criteria}.}
        \label{fig:volume-hist}
    \end{minipage}
\end{figure*}

Using the methods described in Section~\ref{sec:progenitors} we determined whether there is any structure comparable to that of the SSA22 protocluster, in any of the $19$ massive protoclusters in the SMDPL simulation. We found that none of the protoclusters had any sight lines that show evidence for a double-peaked morphology with best-fit parameters similar to those in SSA22, as defined in Equation~\ref{eqn:criteria}.  Figure~\ref{fig:prog-hist} displays the redshift histogram that, out of all sight lines of the 19 protoclusters, shows the greatest similarity with SSA22 as defined by the KS-test $p$-value ($p=1.8\times10^{-4}$). Even this distribution does not show a double-peaked morphology. By observing the spatial distribution of progenitor halos we can understand why there is very little extended structure present.  The range of redshifts present in the SSA22 protocluster corresponds to a spatial separation of $\Delta z = 0.045$ ($\sim40\ h^{-1} \rm \ cMpc$), while the $z\sim3$ halo progenitors of a single massive $z=0$ cluster in the SMDPL simulation typically extend over $\Delta z=0.015$ ($\sim 13\ h^{-1} \rm \ cMpc$).  Sufficiently high peculiar velocities could perturb the redshifts outside the primary structure, however the collapsing nature of these protoclusters tends to compress the redshift distribution on such scales, not expand it.  In summary, comparison with the SMDPL simulation demonstrates that the double-peaked morphology observed in the SSA22 redshift histogram does not comprise the coalescing progenitors of a single $z=0$ structure.

\subsection{Surrounding Volume Halos}
\label{volume-res}

The approach described in the previous section was based on a starting assumption that the entire double-peaked structure in SSA22 corresponds to the progenitor of a single $M\ge10^{15}\ h^{-1}$\msun\ cluster at $z=0$.  Therefore we restricted our analysis to include only the $z\sim3$ progenitor halos of such $z=0$ clusters. Using the alternative approach described in Section~\ref{sec:volume}, we attempt to find structures within the volumes surrounding protoclusters in the simulation at $z\sim3$ that, when ``observed" (as we observe the SSA22 field), produce redshift histograms at $z\sim3$ that are similar to that in SSA22.  We then used the SMDPL simulation to characterize the evolution of such structures to $z=0$.

When examining the distributions of halos in the more extended volumes surrounding massive cluster progenitors, we do find sight lines yielding redshift histograms similar to that of the SSA22 protocluster based on the SMDPL merger trees \citep{Klypin2016, Behroozi2013a}.  Figure~\ref{fig:volume-hist} shows an example of a redshift histogram ($p=0.85$) computed for a single sight line of one simulated protocluster that fits our criteria for similarity to the SSA22 redshift histogram.  In addition to finding sight lines that satisfy our similarity criteria stated in Equation~\ref{eqn:criteria}, we find that many of these `good' sight lines occur from similar viewing angles, suggesting that they are due to real structure, and not statistical flukes. 

We separate the 19 protoclusters into three categories based on the number of distinct sight line groups present in each protocluster. The three categories are: ``no sight lines", ``single sight line group", and ``multiple sight line groups". To assign a protocluster to one of the categories we first looked at the $p$-values distributed throughout the sight-lines.  Figures~\ref{fig:3panel-none}(a)-\ref{fig:3panel-mult}(a) show a projection of the $p$-value distribution as a function of sight-line.  We then looked in detail at the redshift histogram produced when observing along a sight-line with a high $p$-value that passes through a possible cluster progenitor to confirm that it satisfied the similarity criteria of Equation~\ref{eqn:criteria}.  Figure~\ref{fig:3panel-mult}(c) shows an example where the double-peaked structure of the redshift distribution can clearly be seen. If several of these sight-lines are clustered around a specific viewing angle, we consider the viewing angles to be a sight-line group.  Finally, we categorize each protocluster volume based on the number of sight-line groups.   Below we describe the three categories to which we assign each protocluster, with an example from each category detailing the important features in each case.

\subsubsection{No sight lines}
\begin{figure*}
\includegraphics[width=0.95\textwidth]{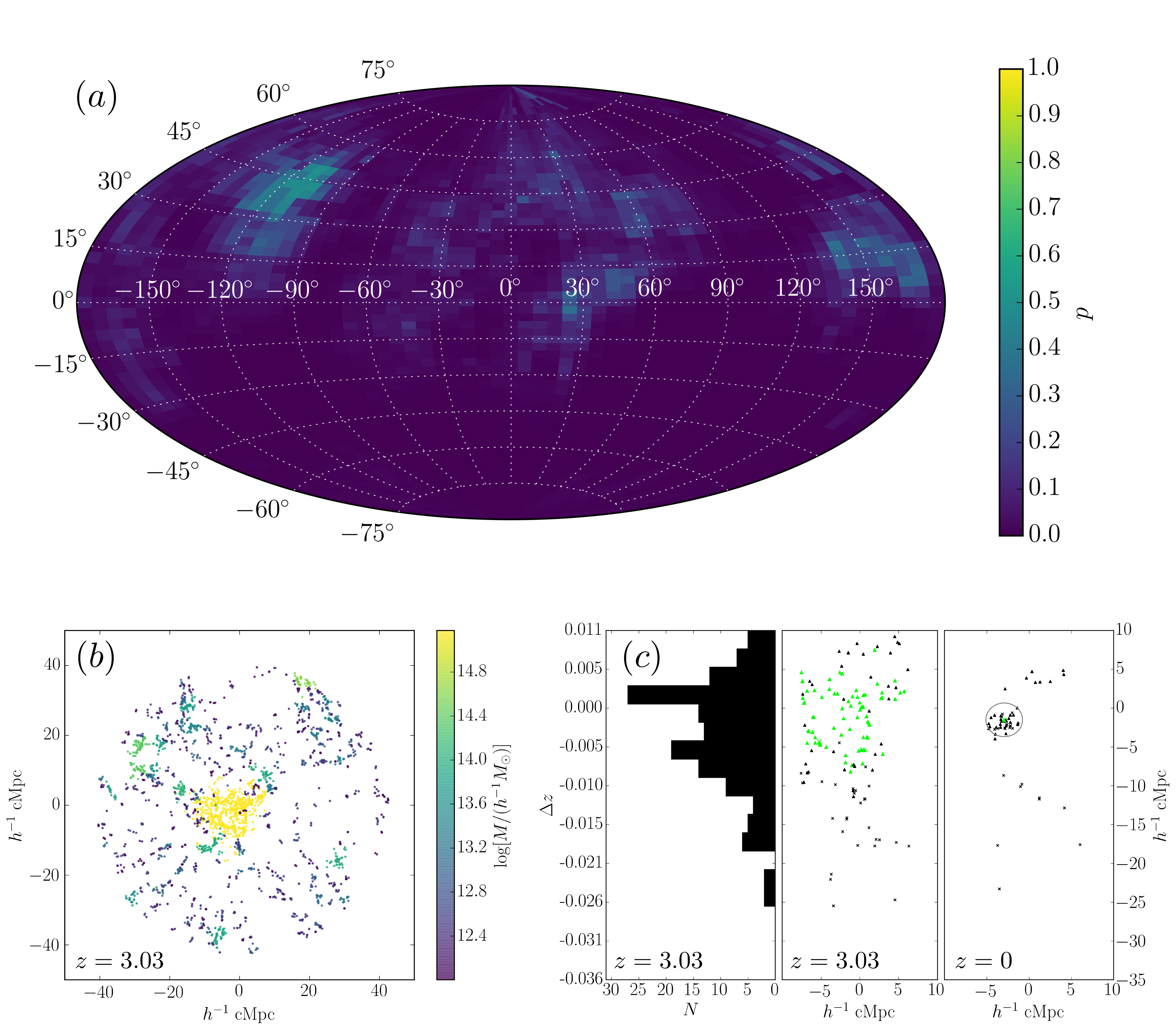}
\caption{Example simulation results for a protocluster volume in the ``no sight lines" category. (a): Mollweide projection of KS $p$-values calculated for redshift histograms created from each sight line toward the protocluster.  Each `pixel' in the projection represents a single sight line.   `Pixels' with higher $p$-values are sight lines that have redshift histograms comparable to that observed in SSA22, however none of the sight lines in category meet our criteria for similarity (Equation~\ref{eqn:criteria}). (b): Scatter plot of halos in the volume surrounding a protocluster, colored by their $z=0$ descendant mass. The yellow points in the center are the $M\sim 10^{15}\ h^{-1}$\msun\ cluster progenitors.  (c): Spatial positions of halos selected from a typical sight line at $z=3.03$ (middle), their corresponding redshift histogram at $z=3.03$ without including sightline-dependent peculiar velocity corrections (left), and their descendant positions at $z=0$ (right).  In the middle panel, $z=3.03$ halos contained in $z=0$  halos with $M>10^{14}\ h^{-1}$\msun\ are colored based on their cluster membership.  The green points at $z=3.03$, which make up the main protocluster, have merged into a single halo at $z=0$. Halos in the right panel with masses $M>10^{14}\ h^{-1}$\msun\ are drawn with their corresponding $R_{200}$ radii.}
\label{fig:3panel-none}
\end{figure*}

One subset of protocluster volumes in the SMDPL simulation that we investigated did not give rise to a double-peaked redshift histogram from any of the sight line viewing angles.  An example of a protocluster in this category is shown in Figure~\ref{fig:3panel-none}.   Figure~\ref{fig:3panel-none}(a) shows the KS $p-$value calculated from the redshift histogram of each sight line.  While there are some sight lines with elevated p-values, there are not multiple adjacent sight lines with elevated $p-$values at any particular viewing angle.  The absence of double peaked histograms in this cluster is expected given the lack of any nearby massive cluster in the SMDPL simulation volume.  This lack of adjacent structure is shown Figure~\ref{fig:3panel-none}(b), where the descendant mass of each halo is displayed.  We find that $2/19=11\%$ protoclusters in our sample fall into this category.

\subsubsection{Single sight line group}  
\begin{figure*}
\includegraphics[width=0.95\textwidth]{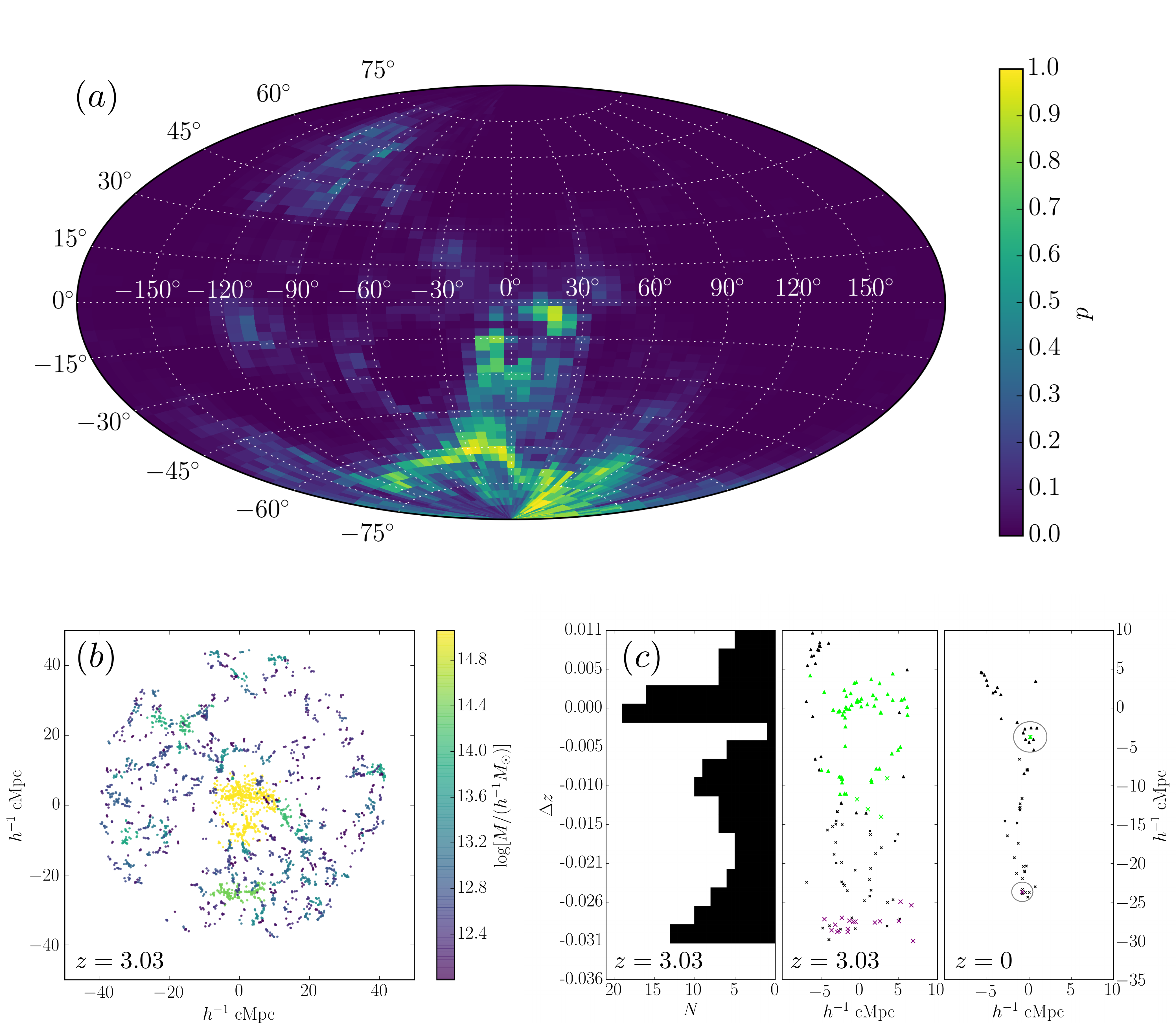}
\caption{Example simulation results for a protocluster volume in the ``single sight line" category. (a): Same as Figure~\ref{fig:3panel-none}(a). The group of sight lines near $\phi=-90$ (south pole) all have similar redshift histograms, suggesting that their double peaks are not due to random variance. The sight lines whose redshift histograms satisfy Equation~\ref{eqn:criteria} are a small subset of the bright `pixels' in this panel. (b): Same as Figure~\ref{fig:3panel-none}(b). A $M>10^{14}\ h^{-1}$\msun\ protocluster can be seen as a collection of green points at $(0,-25)$.  (c): Same as Figure~\ref{fig:3panel-none}(c).  In the middle panel, halos with descendant masses $>10^{14}\ h^{-1}$\msun\ are colored based on their cluster membership.  Points are also displayed as a triangle or a `$\times$' for their membership in the larger or smaller peak respectively determined after adjusting their redshifts due to their peculiar velocities.  Results from one sight line that produced a double-peaked redshift histogram (left) based on the velocities and positions of halos at $z=3.03$ (middle). The two protoclusters that give rise to the double-peaked redshift histogram remain distinct to $z=0$ (right).  At $z=0$ (right) these two groups of halos have each collapsed to a single point.  At the $z=3.03$ epoch, an absence of halos is present at $\Delta z \sim -0.015$ (left, center).  At $z=0$ (right) the two groups of halos have collapsed to form distinct clusters.  Halos with masses $M>10^{14}\ h^{-1}$\msun\ are drawn with their corresponding $R_{200}$ radii.  The two groups are also easily seen in the redshift histogram (left), which does not include corrections based on halo peculiar velocities.}
\label{fig:3panel-one}
\end{figure*}

Another subset of protocluster volumes each yield a single group of closely packed sight lines that produce double-peaked histograms.  Figure~\ref{fig:3panel-one} shows an example of a protocluster in this category.  In this example, many sight lines near the southern pole have high $p$-values suggesting that there is some structure arising in the redshift histograms.  In addition, many of these sight lines also fit our criteria for similarity to the SSA22 redshift histogram, given in Equation~\ref{eqn:criteria}.  The viewing angle of these sight lines is coincident with the progenitor of a second, massive ($M=10^{14.4}\ h^{-1}$\msun) protocluster.   Figure~\ref{fig:3panel-one}(b) shows this protocluster toward the bottom of the panel.  We display the positions of halos from one sight line that shows a similar redshift histogram to that of SSA22 in Figure~\ref{fig:3panel-one}(c).  At $z\approx3$, the main and adjacent structure appear as two separate groups of halos, separated by a lower-density gap. In many cases, the smaller group of halos is the progenitor of a cluster with mass comparable to the expected mass of the blue (smaller) redshift peak in SSA22 at $z=0$ ($M\sim0.7\times10^{15}\ h^{-1}$\msun). Most of the halos that make up the larger and smaller redshift peaks are progenitors of either the main or neighboring cluster.  At $z=0$, the two structures have collapsed into two distinct clusters. We find that $9/19=47\%$ protoclusters in our sample fall into this category.

\subsubsection{Multiple sight line groups}  
\begin{figure*}
\includegraphics[width=0.95\textwidth]{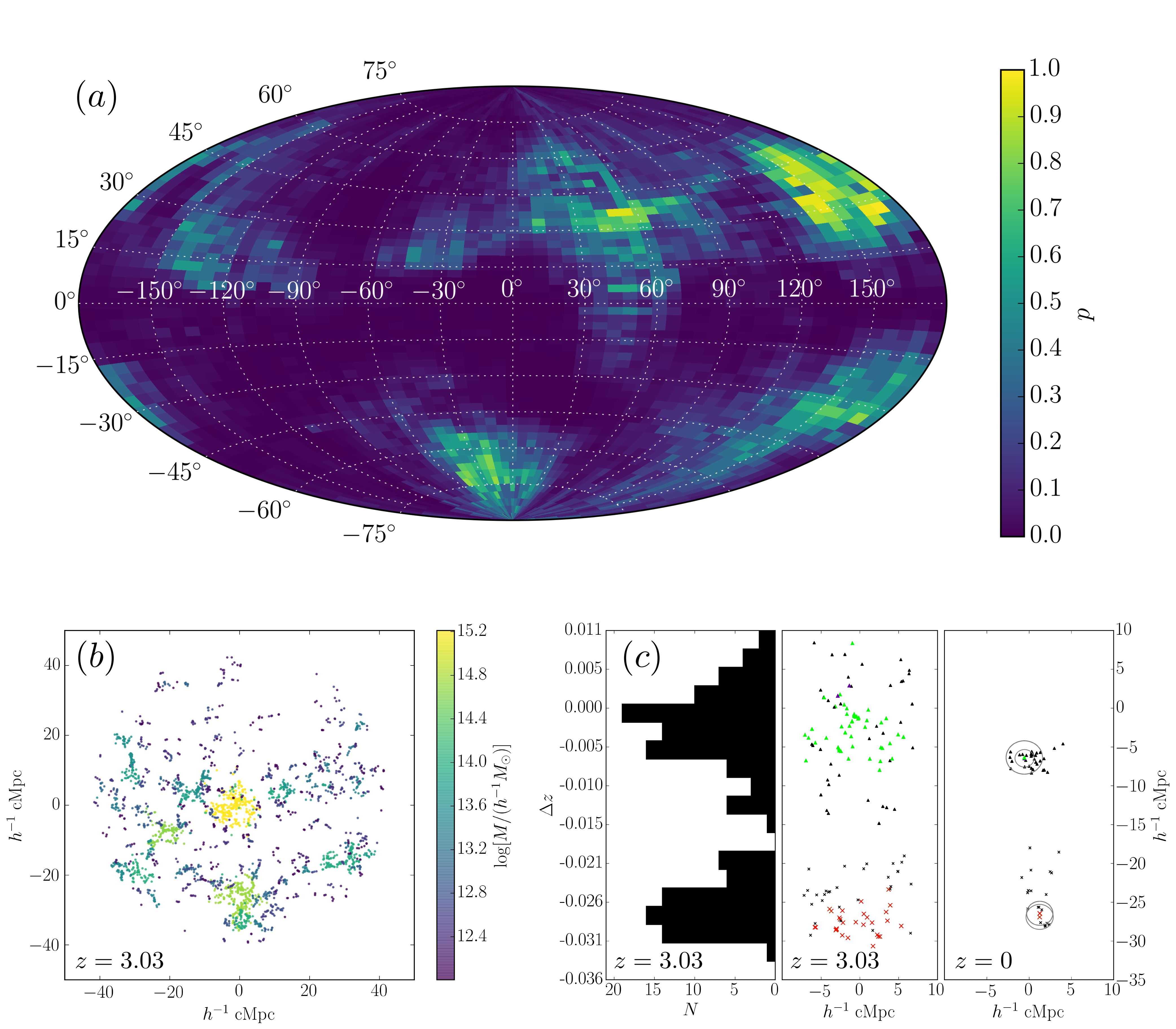}
\caption{Example simulation results for a protocluster volume in the ``multiple sight lines" category. (a): Same as Figure~\ref{fig:3panel-one}(a).  Several distinct groups of sight lines are visible at different viewing angles.  (b): Same as Figure~\ref{fig:3panel-one}(b).  In the volume surrounding this protocluster, multiple other protoclusters can be seen as groups of green points at $(0,-30)$ and $(-20,10)$.  (c): Same as Figure~\ref{fig:3panel-one}(c). At the $z=3.03$ epoch (left, middle), the two distinct groups of halos can be clearly seen at $\Delta z =-0.026$ and $\Delta z = 0.0$.  The halos in each of these two groups have different $z=0$ descendants.  Massive halos at $z=0$ that appear in the same position are separated in the into-the-page direction.}
\label{fig:3panel-mult}
\end{figure*}

The last category consists of protocluster volumes that each contain more than one distinct group of adjacent sight lines. Each of these groups is composed of many closely packed sight lines that produce a double peaked histogram.  Figure~\ref{fig:3panel-mult} shows an example of a protocluster in this category. The KS $p$-value distribution (Figure~\ref{fig:3panel-mult}, a) shows similar properties to the distribution presented in the `single sight line' case.  Protoclusters in this category however, show multiple separate viewing angles comprised of many sight lines with elevated $p$-values, as seen by the different groups of green points.  Each one of these separate viewing angles corresponds to the presence of another nearby massive protocluster. Similar to the adjacent structures in the `single sight line' group, many of the neighboring structures in the `multiple sight line' category have masses comparable to the predicted $z=0$ mass of the blue redshift peak in SSA22.  The centers of these neighboring protoclusters typically lie 10-20 $h^{-1}\rm \ cMpc$ away from the main protocluster.  All neighboring protoclusters are separate from each other at $z\sim3$, and the majority of halos that make up double peaked histograms are members of the main protocluster, and a single neighboring protocluster, as no sight lines intersect multiple neighboring protoclusters.  The neighbors, in addition to the main protocluster, all remain distinct as they collapse to separate structures at $z=0$.  
We find that $8/19=42\%$ of the protoclusters in our sample fall into this category.

\section{Discussion}
\label{sec:discussion}

\subsection{Comparison with Analytic Predictions}
\label{sec:analytic-compare}
We found that the double-peaked redshift histogram of the SSA22 protocluster is the reflection of the presence of less-massive ($>10^{14}\ h^{-1}$\msun) protoclusters in its vicinity. We can use a simple analytic approach to explain quantitatively the prevalence of neighboring, less-massive clusters around $>10^{15}\ h^{-1}$\msun.  Due to halo biasing, we expect the most massive clusters at $z=0$, which lie on an enhanced density peak, to be surrounded by smaller, but still massive, nearby clusters  \citep{Kaiser1984, Barkana2004}.  Using the halo-halo correlation function and the halo mass function, we calculated the number of clusters at a given distance away from some of the most massive clusters.  In this section, in order to more accurately compare to simulations, we adopt a cosmology consistent with the SMDPL simulation: $\Omega_{m} = 0.308$, $\Omega_{\Lambda}=0.692$, $h=0.677$, $\sigma_8=0.8228$, \citep{Planck2013}.

We define the halo-halo correlation as the excess probability of finding a neighbor at a distance $r$ and in the volume $\delta V$ as
\begin{equation}
\delta P = n\delta V (1+\xi(r) ),
\end{equation}
where $n$ is the average number density of halos \citep{Peebles1980}.  We use linear bias to relate the linear matter correlation function, $\xi_{lin}(r)$, to the two-point correlation function of halos with masses $M_1$ and $M_2$, $\xi_{hh}(M_1, M_2, r)$, by
\begin{equation}
\xi_{hh}(M_1, M_2, r) = b(M_1)b(M_2)\xi_{lin}(r).
\end{equation}
To calculate the linear bias factor, $b(M)$, we adopt the definition given by \citet{Quadri2007}:
\begin{equation}
b_h(M) = 1+\frac{1}{\delta_c} \left[  \nu^{\prime 2} + b\nu^{\prime 2(1-c)} - \frac{\nu^{\prime 2c}/\sqrt{a}}{\nu^{\prime 2c}+b(1-c)(1-c/2)} \right],
\end{equation} 
where $\nu^{\prime} = \sqrt{a}\delta_c/\sigma(M,z)$, $\sigma(M,z)$ is the mass variance on scales of $R=\left(\frac{3M}{4\pi \bar{\rho}} \right)^{1/3}\ h^{-1} \rm \ Mpc$, and $\bar{\rho}$ is the mean matter density of the universe.  As in \citet{Quadri2007}, we  use values of $\delta_c = 1.686$, $a=0.707$, $b=0.5$, and $c=0.6$.

We calculate the linear mass correlation function from the power spectrum of fluctuations, $P(k)$, using:
\begin{equation}
\xi_{lin}(R) = \frac{1}{2\pi^2}\int_0^{\infty}P(k) \frac{\sin{kR}}{kR}k^2 dk
\end{equation}
and derive a power spectrum based on the methods described in \citet{Naoz2005}.

Using the halo-halo correlation function, we predicted the mean number of halos, with masses $M\ge M_2$, within a surrounding volume centered on a halo with mass $M_1$ using:
\begin{equation}
\langle N(R) \rangle = n\int_0^R 4\pi r^2 \left[1+\xi_{hh}(r)\right] dr.
\end{equation}
Where $n$ is the average number density of halos, calculated using the halo mass function of \citet{Sheth1999}.\footnote{We obtained the same results when repeating this analysis adopting the halo mass function described in \citet{Tinker2008}, with parameters: $A=0.144$, $a=1.351$, $b=3.113$, and $c=1.187$ provided by \citet{Rodriguez-Puebla2016}.}

\begin{figure}
\includegraphics[width=0.99\linewidth]{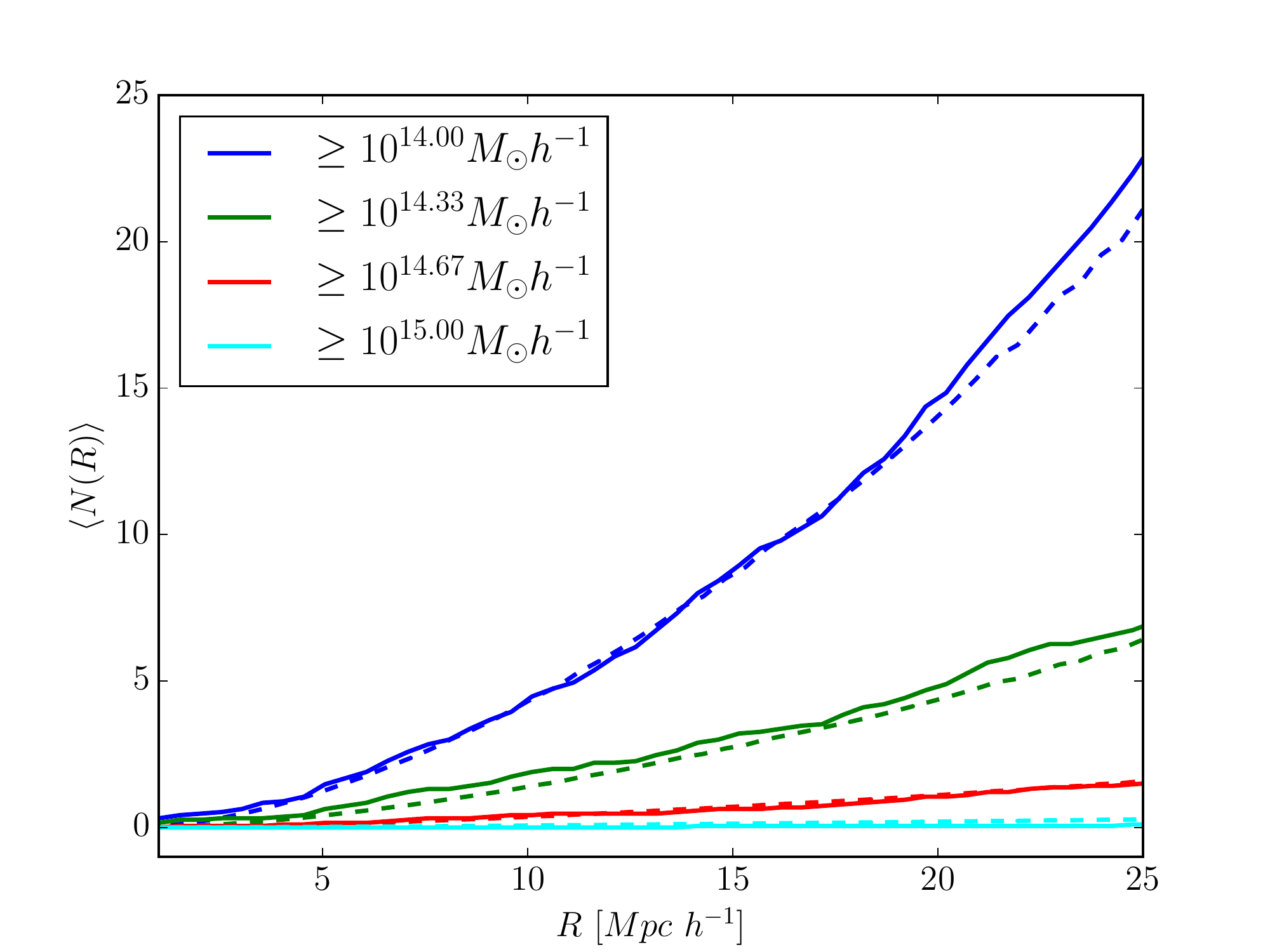}
\caption{Average number of halos of a given mass within a sphere of radius $R$ centered on a $10^{15}h^{-1}$\msun cluster at $z=0$.  Shown here are the analytic predictions (dashed lines) calculated using the method described in Section~\ref{sec:analytic-compare}, compared to the number $N(R)$ measured directly from the SMDPL simulation using $\ge 10^{15}h^{-1}$\msun\ halos as the central halo (solid lines).  
}
\label{fig:analytic}
\end{figure}

Using the method described here, we obtain an analytic prediction for the prevalence of $>10^{14}\ h^{-1}$\msun\ clusters as a function of distance from a $>10^{15}\ h^{-1}$\msun\ cluster at $z=0$.  We then compare our analytic prediction with the results from the SMDPL simulations, and finally with our observations of the SSA22 protocluster.  We calculated the number of halos of a given mass within a given distance, $R$ from the center of a cluster with a mass corresponding to the the mass of one of the 19 clusters present in the SMDPL simulation. This process was repeated for each of the $19$  $M>10^{15}\ h^{-1}$\msun\ clusters in the simulation, and then we averaged the resulting total number. Figure~\ref{fig:analytic} shows this analytic result, calculated using the method described above (dashed lines).  For comparison, Figure~\ref{fig:analytic} displays the number of halos with a given mass and within a given radius, $R$ computed directly from the SMDPL simulation by counting the average number of halos in a sphere with radius $R$ centered on each of the 19 $M>10^{15}\ h^{-1} M_{\odot}$ clusters at $z=0$ (solid lines).

In the case of the SSA22 protocluster, the adjacent structure lies at a distance of $D\approx20\ h^{-1}\rm \ cMpc$ calculated from the difference in peak redshifts neglecting the effects of infall.  Within this distance, our analysis predicts $\sim 1-2$ clusters with a mass comparable to the mass of the blue peak of SSA22. This number increases by $\sim20\%$ when the mass of the central cluster is doubled. This analytic prediction is consistent with our results which place more protoclusters in the ``single sight-line" category, compared to the other categories.  We also predict $\sim10$ clusters with masses $\sim10^{14}\ h^{-1}$\msun\ within this distance, which is again consistent with the simulations.  However, the neighboring clusters that give rise to double-peaked redshift histograms typically have masses of $\ga 3\times 10^{14}\ h^{-1}$\msun.

\subsection{Observing Frequency}
\label{sec:obs-frequency}
We have determined which, if any, sight lines in a given simulated protocluster produce redshift histograms that present a double-peaked morphology, and whether they are similar to the observed redshift histogram in SSA22.  In this section, we discuss the probability of observing a double-peaked redshift histogram, based on our analysis of protoclusters in the SMDPL simulation.   For this analysis, we calculated the density of protoclusters that, when observed, would result in a redshift histogram that contains two peaks similar to that of SSA22, or any structure beyond a single redshift peak.  By searching through sight lines across all protoclusters in the simulation, we determined the frequency at which observations of massive protoclusters would yield double-peaked redshift histograms. 

We started by calculating the covering fraction of sight lines that produced double-peaked redshift histograms.  For an individual protocluster, we calculated the total covering fraction by summing up the contribution from each sight line that we have determined to be double peaked.  The area on the sky covered by a single sight line is given by:
\begin{equation}
\Delta \Omega = \int_{\theta}^{\theta+\Delta \theta} \int_{\phi}^{\phi+\Delta \phi} d\theta d\phi ,
\end{equation}
where $\Delta \theta = 6\deg$, $\Delta \phi = 3\deg$, and $(\theta, \phi)$ is the angle of the sight line.  For a given protocluster, the covering fraction of sight lines with redshift histograms similar to that of SSA22 is:
\begin{equation}
F = \frac{\Omega}{4\pi}
\end{equation}
where $\Omega=\sum \Delta \Omega$ is the total solid angle covered by the relevant sight lines.

On average, the covering fraction of sight lines for a given protocluster is $F=0.025 \pm 0.017$, with values for individual protoclusters ranging from $F=0$ for protoclusters in the ``no sight lines" category, to $F=0.065$ for a protocluster in the ``multiple sight lines" category.  We also consider counting sight lines that are better fit by two peaks, but whose fitting parameters may not fit the criteria presented in Equation~\ref{eqn:criteria}.  Such sight lines contain evidence of structure beyond the main protocluster, but, when observed, do not produce redshift histograms similar to that of SSA22.  The average covering fraction of such additional sight lines is $F=0.13$. To determine the occurrence rate of structures similar to those observed in SSA22, we multiply the covering fraction of sight lines producing double peaked histograms by the number density of massive protoclusters in the SMDPL simulation, $19 / 400^3 \ h^{3}\rm \ cMpc^{-3} = 296\ \mathnormal h^{3}\rm \ Gpc^{-3}$. We therefore calculated the cosmic abundance of observing structure similar to that of the SSA22 protocluster to be $n=7.4\ h^{3}\rm \ Gpc^{-3}$. This density suggests that the observed structure in the SSA22 protocluster is rare, and its discovery unexpected within the $1.07\times10^{-3}\ h^{3} \rm \ Gpc^{-3}$ volume of the survey that discovered it \citep{Steidel2003}.  Even placing a less stringent similarity requirement for the simulated redshift histograms (i.e., some evidence for structure ($p>0.4$), as defined in Section~\ref{sec:progenitors} without strictly satisfying Equation~\ref{eqn:criteria}), we find a cosmic abundance of only $n=38\ h^{3}\rm \ Gpc^{-3}$, which still makes the discovery of SSA22 extremely fortuitous within the LBG survey volume. Hints of bimodality have been seen in other protoclusters \citep[e.g.;][]{Kuiper2011, Venemans2007}. However, better spectroscopic sampling as well as evidence of a spatial offset between redshift peaks are required to determine the similarity of these structures to the observed large-scale structure in SSA22.

\section{Summary and Conclusion}
We have used an updated spectroscopic sample to measure the overdensity and mass of the SSA22 protocluster, and its associated structure.  We then attempted to understand these results using the SMDPL cosmological simulation, and a simple analytic approach.  In detail:

\begin{enumerate}
\item We used an updated sample of spectroscopic redshifts of LBGs in the SSA22 field to measure the overdensities of the total SSA22 region ($\delta_{t,gal}=7.6\pm 1.4$), and the blue and red peaks present in its redshift histogram ($\delta_{b,gal}=4.8 \pm 1.8$, $\delta_{r,gal}=9.5 \pm 2.0$).  We utilized updated overdensity measurements to calculate the masses of the total region ($M_t=(3.19 \pm 0.40) \times  10^{15} h^{-1}
M_{\odot}$), the blue redshift peak ($M_b = (0.76 \pm 0.17) \times  10^{15} h^{-1}
M_{\odot}$), and red redshift peak ($M_r = (2.15 \pm 0.32) \times  10^{15} h^{-1}
M_{\odot}$).

\item Using our updated predictions for the masses of these two peaks, we made use of the Small MultiDark Planck simulation to determine the nature of the double-peaked redshift distribution.  First, we tested the scenario that the structure in SSA22 is all contained in the progenitor of a single massive cluster.  For this analysis, we looked in the simulation only at halos that would eventually collapse into a single massive ($M>10^{15}\ h^{-1} M_{\odot}$) structure at $z=0$.  From these we created simulated redshift histograms and compared their morphology to the observed redshift distribution in SSA22.  In the $19$ $M>10^{15}\ h^{-1} M_{\odot}$ protoclusters in the simulation that we observed, none had progenitor halo distributions that alone produced a redshift histogram consistent with the double-peaked shape observed in SSA22.  

\item We performed a complementary approach that considered all halos within a certain distance of each individual $M>10^{15}\ h^{-1} M_{\odot}$ protocluster in the simulation, regardless of membership in the associated descendant cluster at $z=0$.  Using this method, we found that $17/19$ of the simulated protoclusters had configurations that, when observed from at least some lines of sight, produced redshift histograms similar to that of the observed distribution in SSA22.  For each of these 17 protoclusters, the viewing angles that produced the matching redshift histograms contain the main overdensity along with a neighboring aligned, but less massive, overdensity.  Following these adjacent protoclusters through time in the simulation, we saw that the two structures in the volume remained distinct to $z=0$, demonstrating that the second peak in the redshift histogram can be caused by a separate virialized structure from the main protocluster. 
 
\item We further investigated the results from the simulation using a simple analytic approach.  Using the halo-halo correlation function derived from the dark matter power spectrum, we predicted the number of halos of a given mass within a distance $R$ from a massive cluster.  The results from this analysis are consistent with what we have seen in the simulation, predicting $\sim1-2$ massive halos surrounding each main cluster capable of producing a second peak in the redshift distribution. 

\item Finally, using the covering fraction of sight lines of simulated protoclusters that produced double-peaked redshift histograms, and the number density of massive protoclusters, we predicted the occurrence of a structure similar to that observed in SSA22 to be $7.4\ h^3 \rm \ Gpc^{-3}$.
\end{enumerate}

Previous estimates of the mass of the SSA22 overdensity have been produced by considering the volume containing the red and blue peak as a single massive protocluster.  By treating the entire region as a single overdensity, previous studies have overestimated the mass of the main, $M\sim10^{15}\ h^{-1} M_{\odot}$ protocluster. The existence of the second (blue) peak must be considered in order to obtain an accurate measurement of the mass.

Due to the limited area that our observations cover, we are restricted to observing structure coincident with the line-of-sight to the main protocluster. In order to fully understand the connection between the structure and the main protocluster, deep and densely sampled spectroscopic observations must be performed in an area extending at least $\sim20 \ h^{-1} \rm \ Mpc$ ($\sim11'$) away from the center of the protocluster.  This approach would allow us to not only fully map the structure already observed, but also find other massive nearby structures, if present.  In addition to wider-field observations of the SSA22 protocluster, an in-depth analysis of the structure present in additional known protoclusters \citep[e.g., HS1700+643 at $z=2.299$, and HS1549+195 at $z=2.842$;][]{Steidel2005, Steidel2011} found in larger cosmic volumes will demonstrate the variety of environments of the most massive structures in the universe as they formed.

\section*{Acknowledgments}
The authors gratefully acknowledge the Gauss Centre for Supercomputing e.V. (www.gauss-centre.eu) and the Partnership for Advanced Supercomputing in Europe (PRACE, www.prace-ri.eu) for funding the MultiDark simulation project by providing computing time on the GCS Supercomputer SuperMUC at Leibniz Supercomputing Centre (LRZ, www.lrz.de).
The Bolshoi simulations have been performed within the Bolshoi project of the University of California High-Performance AstroComputing Center (UC-HiPACC) and were run at the NASA Ames Research Center.
We wish to extend special thanks to those of Hawaiian ancestry on whose sacred mountain we are privileged to be guests. Without their generous hospitality, most of the observations presented herein would not have been possible.
SN acknowledges partial support from a Sloan Foundation Fellowship. C.C.S. acknowledges support from NSF grants AST-0908805 and AST-1313472.

\clearpage

\clearpage
\end{document}